\documentclass[journal]{IEEEtran}

\usepackage{pstricks}
\usepackage{graphicx}
\usepackage{psfrag}

\usepackage[cmex10]{amsmath}
\usepackage{amssymb}

\usepackage{color}

\graphicspath{{./pics/}}

\def\ve#1{{\mathchoice{\mbox{\boldmath$\displaystyle #1$}}%
              {\mbox{\boldmath$\textstyle #1$}}%
              {\mbox{\boldmath$\scriptstyle #1$}}%
              {\mbox{\boldmath$\scriptscriptstyle #1$}}}}
\DeclareSymbolFont{AMSb}{U}{msb}{m}{n}
\DeclareSymbolFontAlphabet{\mathbb}{AMSb}

\def\F{\mathbb{F}}

\def\conc{\otimes}
\def\concb{\odot}
\def\labeling{\mathcal L}
\def\mean{M}

\newcommand{\W}{\mathsf{W}}
\newcommand{\B}{\mathsf{B}}

\newcommand {\Es} {E_{\mathrm{s}}}            
\def\dB{\mathrm{dB}}

\def\No{N_0}

\begin{document}  
\title{Polar-Coded Modulation}
\author{Mathis~Seidl,~Andreas~Schenk,~Clemens~Stierstorfer, and~Johannes~B.~Huber%
\thanks{All authors are
  with the Institute of Information Transmission,
  Friedrich-Alexander-Universit{\"a}t Erlangen-N{\"u}rnberg, 91058 Erlangen,
  Germany.}
\thanks{E-mail: {\tt\{seidl,schenk,clemens,jbhuber\}}@LNT.de.} 
\thanks{Parts of this paper have been submitted for presentation at IEEE ISIT '13.}
}
\maketitle
%
\begin{abstract}
A framework is proposed that allows for a joint description and optimization of both 
binary polar coding and $2^m$-ary digital pulse-amplitude modulation (PAM) schemes such 
as multilevel coding (MLC) and bit-interleaved coded modulation (BICM). 
The conceptual equivalence of polar coding and multilevel coding is pointed out in detail. 
Based on a novel characterization of the channel polarization phenomenon, rules for 
the optimal choice of the labeling in coded modulation schemes employing polar 
codes are developed. Simulation results regarding the error performance of the proposed 
schemes on the AWGN channel are included. 
\end{abstract}
\section{Introduction}
\label{sec:intro}
\noindent
Polar codes~\cite{Arikan:09} are known as a low-complexity binary coding scheme that provably approaches the capacity of arbitrary symmetric binary-input discrete memoryless channels (B-DMCs). 
The generalization to $M$-ary channels ($M>2$) has been the subject of various works, cf., e.g. \cite{Sasoglu:09, Sahebi:11, Park:12}.
However, the topic of \emph{polar-coded modulation}, i.e., the combination of $M=2^m$-ary digital modulation, especially digital PAM (i.e., ASK, PSK, QAM), and binary polar codes 
for increased spectral efficiency, has hardly been addressed so far. 
In \cite{ShinLimYang:12}, a transmission scheme for polar codes with bit-interleaved coded modulation (BICM)~\cite{CaireTB:98,FabregasMC:now2009} has been proposed, focussing on the interleaver design. 

In this paper, we discuss both the multilevel coding (MLC) construction~\cite{ImaiH:1977, WachsmannFH:99} and BICM. We restrict our considerations to memoryless channels like the AWGN channel (no 
fading). In case of BICM, we follow an alternative approach that differs from~\cite{ShinLimYang:12}. 

It has been observed (cf., e.g.,~\cite{ArikanISIT:11}) that the MLC approach is closely related to that of polar coding on a conceptual level. 
Based on these similarities, we propose a framework that allows us to completely describe both polar coding and $2^m$-ary PAM modulation in a unified context. 
To this end, we introduce so-called channel partitions.
These transformations split an arbitrary memoryless $2^m$-ary channel (e.g., the equivalent-baseband PAM channel in case of PAM modulation) into
$m$ binary-input memoryless channels (so-called \emph{bit channels}). 

We distinguish two classes of such binary partitions, sequential and parallel binary partitions.
For the latter, the resulting bit channels are independent. It is thus applicable, e.g., to describe BICM.
For sequential binary partitions, the bit channels depend on each other in a well-defined order -- this class can be used for representing MLC. 
We show that both binary polar coding as well as polar-coded modulation may be described by the concatenation of binary partitions. 

Considering the trade-off between power efficiency and spectral efficiency, this unified description makes it possible to design optimized constellation-dependent 
coding schemes both for MLC and BICM. 

Additionally,  we provide an efficient method for a numerical evaluation of the performance of polar-coded modulation and present extensive 
numerical results for various settings. Using this method, we present a comprehensive comparison of polar-coded modulation based on MLC as well as on BICM, 
and we show results of a comparison to LDPC-coded modulation (for the latter only the common BICM approach is considered).

The paper is organized as follows: 
In Sec.~\ref{sec:ch_transf}, the framework for a joint description of polar coding and $2^m$-ary PAM modulation is developed. 
This framework is then used for describing the polar coding construction in Sec.~\ref{sec:polarcodes}, leading to a novel interpretation of the polarization phenomenon. 
The optimal combination of binary polar coding and $2^m$-ary modulation is discussed in Sec.~\ref{sec:polar_mlc} for the multilevel coding approach and in Sec.~\ref{sec:polar_bicm} 
for bit-interleaved coded modulation (BICM), followed by simulation results for the AWGN channel in Sec.~\ref{sec:sim}.


\section{Channel Transforms}
\label{sec:ch_transf}
\noindent
\subsection{Sequential Binary Partitions}
Let $\W : \mathcal X \rightarrow \mathcal Y$ be a discrete, memoryless channel (DMC) with input symbols $x\in \mathcal X$ (alphabet size $| \mathcal X|= 2^k$), output symbols 
$y\in \mathcal Y$ from an arbitrary alphabet $\mathcal Y$, 
and mutual information $I(X;Y)$.
\footnote{A short remark on the notation: channels are denoted by sans-serif fonts, capital roman letters stand for random variables while boldfaced symbols denote vectors or matrices.}
We define an order-$k$ \emph{sequential binary partition} ($k$-SBP) $\varphi$ of $\W$ to be a channel transform
\begin{equation}
 \varphi : \W \rightarrow \{\B_\varphi^{(0)},\ldots, \B_\varphi^{(k-1)} \}
\end{equation}
that maps $\W$ to an \emph{ordered} set of $k$ binary-input 
DMCs (B-DMCs) which we refer to as \emph{bit channels}. 
For any given $\W$, such a $k$-SBP is characterized  by  a binary labeling rule $\labeling_\varphi$ that maps binary $k$-tuples bijectively to 
the $2^k$ input symbols $x \in \mathcal X$:
\begin{equation}
 \labeling_\varphi :\ [b_0,b_1,\ldots ,b_{k-1}] \in \{0,1\}^k \mapsto x \in \mathcal X\; .
\end{equation}
The number of possible labelings equals $(2^k!)$.

Each bit channel $\B_\varphi^{(i)}$ ($0 \leq i < k$) of a $k$-SBP is supposed to have knowledge of the output of $\W$ as well as of the values transmitted over the bit channels of smaller 
indices $\B_\varphi^{(0)},\ldots ,\B_\varphi^{(i-1)}$. Thus, we have
\begin{equation}
 \B_\varphi^{(i)}    : \{0,1\} \rightarrow \mathcal Y \times \{0,1\}^i \; .
\end{equation}
The mutual information between channel input and output of $\B_\varphi^{(i)}$ assuming equiprobable input symbols is therefore given by 
\begin{equation}
 I(\B_\varphi^{(i)}) := I(B_i ; Y | B_0 , \ldots , B_{i-1})
\end{equation}
which we refer to as the \emph{(symmetric) bit channel capacity} of $\B_\varphi^{(i)}$. 
(If $\W$ is a symmetric channel, this value in fact equals the channel capacity.)
The mutual information of $\W$ is preserved under the transform $\varphi$, i.e.,
\begin{equation}\label{chain_rule}
 \sum_{i=0}^{k-1}I(\B_\varphi^{(i)}) = I(X;Y)
\end{equation}
which directly follows from the chain rule of mutual information~\cite{LIT_itw96, CoverT:91}.

Considering polar-coded modulation, we show that the code construction can be described by SBPs. We are particularly interested in two properties of SBPs, namely the mean value 
and the variance of the bit channel capacities, defined respectively as
\begin{align}
 \mean_\varphi(\W) &:= \frac{1}{k}\sum_{i=0}^{k-1} I(\B_\varphi^{(i)}) = \frac{1}{k} I(X;Y)\label{def_mean}\\ 
 V_\varphi(\W) &:= \frac{1}{k}\sum_{i=0}^{k-1} I(\B_\varphi^{(i)})^2 - \mean_\varphi(\W)^2 \; .\label{def_variance}
\end{align}
Clearly, from \eqref{chain_rule} the mean value $\mean_\varphi(\W)$ in fact depends only on the channel $\W$, rather than on the particular transform $\varphi$. 
It represents the average (symmetric) capacity of $\W$ per transmitted binary symbol. Obviously, 
\begin{equation}
 0 \leq \mean_\varphi(\W) \leq 1
\end{equation}
holds for any DMC $\W$ and any SBP $\varphi$. 
The variance of an SBP $\varphi$ is upper-bounded by 
\begin{equation}
 V_\varphi(\W) \leq \mean_\varphi(\W) (1- \mean_\varphi(\W))
\end{equation}
with equality only iff all $I(\B_\varphi^{(i)})$ are either $0$ or $1$. 
This follows from
\begin{align}\label{max_var}
 V_\varphi(\W) &=   \frac{1}{k}\sum_{i=0}^{k-1} I(\B_\varphi^{(i)})^2 - \mean_\varphi(\W)^2 \\\nonumber
               &\leq \frac{1}{k}\sum_{i=0}^{k-1} I(\B_\varphi^{(i)}) - \mean_\varphi(\W)^2  \\\nonumber
               &= \mean_\varphi(\W) (1- \mean_\varphi(\W)) \; .
\end{align}
and $0 \leq I(\B_\varphi^{(i)}) \leq 1$ for all $0 \leq i < k$.
Note that this upper bound does not depend on the particular labeling $\labeling_\varphi$ but only on the channel $\W$.

An important subset of $k$-SBPs is formed by those transforms whose labeling rules are described by binary bijective linear mappings. 
Let $\W = (\B_0 \times \ldots \times \B_{k-1})$ be a vector channel of $k$ independent B-DMCs $\B_0, \ldots , \B_{k-1}$. 
Then, we call the $k$-SBP
\begin{equation}
 \varphi: \quad  (\B_0 \times \ldots \times \B_{k-1}) \rightarrow \{\B_\varphi^{(0)},\ldots, \B_\varphi^{(k_1-1)} \}
\end{equation}
a \emph{linear $k$-SBP} if its labeling rule is given by
\begin{equation}
 \labeling_\varphi :\ \ve{b} \in \F_2^k \mapsto \ve{b} \cdot \ve{A}_{\varphi} \in \F_2^k \; .
\end{equation}
with $\ve{b} := [b_0,b_1,\ldots ,b_{k-1}]$ and $\ve{A}_{\varphi}$ being an invertible binary $(k,k)$ matrix. 
Clearly, the number of possible linear $k$-SBPs equals the number of non-singular binary $(k,k)$ matrices and is significantly smaller than that of 
general $k$-SBPs.

\subsection{Product Concatenation of SBPs}
Under certain conditions discussed below, it is possible to concatenate two (or more) SBPs in a product form. Let
\begin{equation}
 \varphi: \quad \W \rightarrow \{\B_\varphi^{(0)},\ldots, \B_\varphi^{(k_1-1)} \}
\end{equation}
be an arbitrary $k_1$-SBP and 
\begin{equation}
 \psi: \quad (\B_0 \times \ldots \times \B_{k_2-1}) \rightarrow \{\B_\psi^{(0)},\ldots, \B_\psi^{(k_2-1)} \}
\end{equation}
a $k_2$-SBP that takes a vector channel of $k_2$ independent B-DMCs $\B_0, \ldots , \B_{k_2-1}$ as an input. 
Each of the vector channels $(\B_\varphi^{(i)})^{k_2}$ -- obtained by taking $k_2$ independent instances of $\B_\varphi^{(i)}$ -- can be partitioned by $\psi$.
\begin{figure}
\psfrag{chW}[Bc][Bc][0.6]{$\W$}
\psfrag{Tphi}[Bc][Bc][0.8]{$\varphi$}
\psfrag{Tpsi}[Bc][Bc][0.8]{$\psi$}
\psfrag{B10}[Bc][Bc][0.6]{${\color{blue}\B_{\varphi}^{(0)}}$}
\psfrag{B11}[Bc][Bc][0.6]{${\color{red}\B_{\varphi}^{(1)}}$}
\psfrag{B20}[Bc][Bc][0.6]{$\B_{\varphi \circ \psi}^{(0)}$}
\psfrag{B21}[Bc][Bc][0.6]{$\B_{\varphi \circ \psi}^{(1)}$}
\psfrag{B22}[Bc][Bc][0.6]{$\B_{\varphi \circ \psi}^{(2)}$}
\psfrag{B23}[Bc][Bc][0.6]{$\B_{\varphi \circ \psi}^{(3)}$}
\centerline{\includegraphics[width=0.48\textwidth]{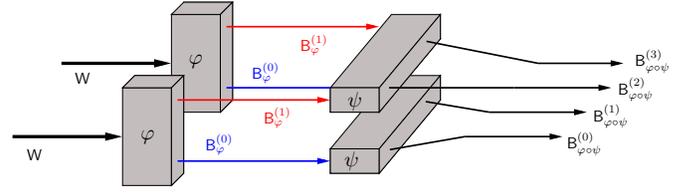}}
\caption{\label{SBP_conc}Concatenation of two $2$-SBPs $\varphi: \W \rightarrow \{\B_{\varphi}^{(0)},\B_{\varphi}^{(1)}\}$ and $\psi: \B^2 \rightarrow \{\B_{\psi}^{(0)},\B_{\psi}^{(1)}\}$.}
\end{figure}
Thus, $\varphi$ and $\psi$ may be concatenated by considering the vector channel $\W^{k_2}$, leading to a product SBP of order $k_1k_2$:
\begin{equation}\label{def_SBP_conc}
 \varphi \conc \psi: \W^{k_2} \rightarrow \{\B_{\varphi\conc\psi}^{(0)},\ldots, \B_{\varphi\conc\psi}^{(k_1k_2-1)} \} \; .
\end{equation}
Here, the bit channels of $\varphi \conc \psi$ are given by
\begin{equation}
 \B_{\varphi \conc \psi}^{(k_2i+j)} : \{0,1\} \rightarrow \mathcal Y^{k_2} \times \{0,1\}^{k_2i+j}
\end{equation}
with symmetric capacities
\begin{equation}
 I(\B_{\varphi \conc \psi}^{(k_2i+j)}) = I(B_{k_2i+j} ; Y_0,\ldots ,Y_{k_2-1} | B_0 , \ldots , B_{k_2i+j-1})
\end{equation}
such that
\begin{equation}\label{mean_conc}
 \frac{1}{k_2}\sum_{j=0}^{k_2-1} I(\B_{\varphi \conc \psi}^{(k_2i+j)}) = I(\B_\varphi^{(i)})
\end{equation}
for all $0\leq i < k_1$ and $0\leq j < k_2$. We remark that the product transform $\varphi \conc \psi$ is completely determined in a unique way
by the individual SBPs $\varphi$ and $\psi$ since their bit channels imply a fixed order. 
Fig. \ref{SBP_conc} shows a simple example of such a product concatenation of two $2$-SBPs that results in a $4$-SBP. 

The product concatenation of SBPs does not influence the mean value of the bit channel capacities, since
\begin{align}\label{mean_invar}
 \mean_{\varphi \conc \psi}(\W^{k_2}) &= \frac{1}{k_1k_2} \sum_{i=0}^{k_1-1} \sum_{j=0}^{k_2-1} I(\B_{\varphi \conc \psi}^{(k_2i+j)})\\\nonumber
                            &= \frac{1}{k_1} I(X;Y) = \mean_{\varphi}(\W) \; .
\end{align}
holds due to the chain rule of mutual information. 
However, the variance of the bit channel capacities increases. It is given by the sum of the variance of the first transform and the averaged variance of the second transform around 
the bit channel capacities of the first one:
\begin{equation}\label{var_conc}
 V_{\varphi \conc \psi}(\W^{k_2}) = V_{\varphi}(\W) + \frac{1}{k_1} \sum_{i=0}^{k_1-1} V_\psi(\B_\varphi^{(i)}) \; .
\end{equation}
This relation is proven in appendix A. 

If $\varphi$ and $\psi$ are linear SBPs with labeling rules specified by $\ve{A}_\varphi$ and $\ve{A}_\psi$, respectively, then their product $\varphi \conc \psi$ is again a linear 
$k_1k_2$-SBP with labeling rule
\begin{equation}\label{SBP_conc_lin}
 \labeling_{\varphi \conc \psi} :\ \ve{b} \in \F_2^{k_1k_2} \mapsto \ve{b} \cdot \ve{P}_{k_1,k_2} \cdot \left( \ve{A}_{\psi} \otimes \ve{A}_{\varphi} \right) \; .
\end{equation}
Here, $ \ve{A}_{\psi} \otimes \ve{A}_{\varphi}$ denotes the Kronecker product of $\ve{A}_{\psi}$ and $\ve{A}_{\varphi}$. 
$\ve{P}_{k_1,k_2}$ is the $(k_1k_2,k_1k_2)$ permutation matrix that maps the ($k_2i+j$)-th component of the vector $\ve{b}$ to position $i+k_1j$ 
(for all $0\leq i < k_1$, $0 \leq j < k_2$).

\subsection{Parallel Binary Partitions}
Let $\W$ be a DMC with $2^k$-ary input as above. In analogy to the sequential approach of SBPs, we define an order-$k$ \emph{parallel binary partition} ($k$-PBP) of $\W$ as a channel transform
\begin{equation}\label{def_PBP}
 \bar \varphi : \W \rightarrow \{\B_{\bar \varphi}^{(0)},\ldots, \B_{\bar \varphi}^{(k-1)} \}
\end{equation}
that maps $\W$ to a set of \emph{independent} B-DMCs. The bit channels of the PBP $\bar\varphi$ are characterized by 
\begin{equation}
 \B_{\bar\varphi}^{(i)}: \{0,1\} \rightarrow \mathcal Y
\end{equation}
with symmetric capacities
\begin{equation}
 I(\B_{\bar\varphi}^{(i)}) := I(B_i;Y)
\end{equation}
for $0 \leq i < k$. Note that for a given $\W$ a $k$-SBP $\varphi$ turns into a $k$-PBP $\bar\varphi$ if the order of the bit channels and, by this, 
the information transfer from bit channels of lower indices are discarded. We refer to $\varphi$ and $\bar\varphi$, 
that share the same labeling rule $\labeling_\varphi$, 
as \emph{corresponding channel partitions}.

Mean value $\mean_{\bar\varphi}(\W)$ and variance $V_{\bar\varphi}(\W)$ are defined in analogy to \eqref{def_mean} and \eqref{def_variance}; however, here the mean value depends on the 
specific PBP $\bar\varphi$ and is (in general) smaller than that of the corresponding SBP $\varphi$
\begin{equation}
 \mean_{\bar\varphi}(\W) \leq \mean_{\varphi}(\W)
\end{equation}
since obviously 
\begin{align}\label{eq:PDvsMSD}
 I(\B_{\bar\varphi}^{(i)}) &= I(B_i;Y)\\\nonumber
                           &\leq I(B_i; Y | B_0,\ldots, B_{i-1}) = I(\B_{\varphi}^{(i)})
\end{align}
holds for all pairs of bit channels. Unfortunately, a general comparative statement on the variances of $\varphi$ and $\bar\varphi$ is not possible due to the 
labeling-dependent mean value $\mean_{\bar\varphi}(\W)$. 

\subsection{Concatenation of PBPs}
In contrast to the case of SBPs, there is no unique way to concatenate parallel binary partitions since the output bit channels $\B_{\bar\varphi}^{(i)}$ of a PBP $\bar\varphi$ are 
mutually independent, allowing for arbitrary permutations between the particular PBPs.

However, we point out that the (unpermuted) concatenation of a $k$-PBP $\bar\varphi$ as in \eqref{def_PBP} with a $k$-SBP $\psi$ (that accepts $k$ B-DMCs as an input), i.e., 
\begin{equation}
 \bar \varphi \concb \psi : \W \rightarrow \{\B_{\bar \varphi \concb \psi}^{(0)},\ldots, \B_{\bar \varphi \concb \psi}^{(k-1)} \} \; ,
\end{equation}
that simply connects the (independent) output channels of $\bar \varphi$ to the input of $\psi$, 
results in sort of a ``degraded $k$-SBP'' with labeling rule $\labeling_{\varphi\concb\psi}$ in the sense 
that its bit channels imply a fixed order while their capacities do not sum up to $I(X;Y)$.
The bit channels of this transform are given by
\begin{equation}
 \B_{\bar\varphi \concb \psi}^{(i)}: \{0,1\} \rightarrow \mathcal Y \times \{0,1\}^i
\end{equation}
($0\leq i< k$) with symmetric capacities
\begin{equation}
 I(\B_{\bar\varphi \concb \psi}^{(i)}) := I(B_i;Y|B_{\bar\varphi,0},\ldots , B_{\bar\varphi,i-1})
\end{equation}
where $B_{\bar\varphi,i}$ ($0\leq i< k$) denote the labels at the output of $\bar\varphi$.
The sum of bit channel capacities equals the value from the PBP $\bar\varphi$:
\begin{equation}
 \frac{1}{k}\sum_{i=0}^{k-1}I(\B_{\bar\varphi \concb \psi}^{(i)}) = \mean_{\bar\varphi}(\W) \leq \mean_{\varphi}(\W) = \frac{1}{k}I(X;Y)\; ;
\end{equation}
thus, the transform $\bar\varphi\concb\psi$ is in general not a SBP. 

In case that $\bar \varphi$ and $\psi$ are linear channel transforms represented by $\ve{A}_\varphi$ and $\ve{A}_\psi$, respectively, the concatenation $\bar \varphi \concb \psi$ 
is again a linear transform characterized by the labeling rule
\begin{equation}
 \labeling_{\varphi \conc \psi} :\ \ve{b} \in \F_2^{k} \mapsto \ve{b} \cdot \left( \ve{A}_{\psi} \cdot \ve{A}_{\varphi} \right) \; ,
\end{equation}
i.e., the common matrix product of $\ve{A}_\psi$ and $\ve{A}_\varphi$.


\section{Polar Codes}
\label{sec:polarcodes}
\noindent

Polar codes, as introduced by Ar{\i}kan~\cite{Arikan:09}, have been shown to be a channel coding construction that provably achieves the symmetric capacity of arbitrary binary-input discrete memoryless channels (B-DMCs) under low-complexity encoding and successive cancellation (SC) decoding. 
For sake of simplicity, we focus on Ar{\i}kan's original construction in this paper; the generalization to polar codes based on different kernels (as considered, e.g., in~\cite{Korada:10}) 
is straightforward. Furthermore, we restrict our considerations to the SC decoding algorithm as in~\cite{Arikan:09}; though, our results regarding the code construction are also 
valid for other (better performing) decoders that are based on the SC algorithm, as, e.g., list decoding~\cite{TalVardy:11}.

\subsection{Code Construction}

Let $\B : \{0,1\} \rightarrow \mathcal Y$ be a B-DMC and $I(\B)$ its symmetric capacity, i.e., the mutual information of $\B$ assuming equiprobable binary input symbols. 
Encoding takes place in the binary field $\F_2$. 
The encoding operation for a polar code of length $N$ may be 
described by multiplication of a binary length-$N$ vector $\ve{u}$ -- containing the information symbols as well as some symbols with fixed values (so-called \emph{frozen symbols}) 
that do not hold any information -- 
with a generator matrix $\ve{G}_N$ that is defined by the 
recursive relation
\begin{equation}
 \ve{G}_N = \ve{B}_N\ve{F}_N\; , \quad \ve{F}_{2N} = \ve{F}_2 \otimes \ve{F}_N \; , \quad  \ve{F}_2 = \left[ \begin{matrix} 1 & 0\\ 1 & 1 \end{matrix} \right]  \label{eq:polarc:G}
\end{equation}
where $N$ is a power of two and $\otimes$ again denotes the Kronecker product. $\ve{B}_N$ denotes the $(N,N)$ bit-reversal permutation matrix~\cite{Arikan:09}. 
The resulting codeword $\ve{c}=\ve{u}\ve{G}_N$ is then transmitted in $N$ time steps over the binary channel $\B$. 
\psfrag{Wchan}[Bc][][0.8]{$\B$}
\psfrag{galp2}[Bc][Bc][0.9]{$\oplus$}
\psfrag{u1}[Bc][Bc][0.8]{$u_0$}
\psfrag{u2}[Bc][Bc][0.8]{$u_1$}
\psfrag{x1}[Bc][Bc][0.8]{$u_0 \oplus u_1$}
\psfrag{x2}[Bc][Bc][0.8]{$u_1$}
\psfrag{yps1}[Bc][Bc][0.8]{$y_0$}
\psfrag{yps2}[Bc][Bc][0.8]{$y_1$}
\begin{figure}
\centerline{\includegraphics[width=0.4\textwidth]{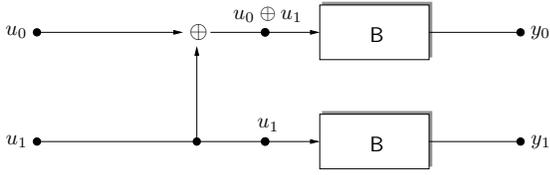}}
\caption{\label{Polar_2}Polar coding construction for $N=2$.}
\end{figure}

The code construction is based on a channel combining and channel splitting operation~\cite{Arikan:09} that may be represented as a linear $2$-SBP
\begin{equation}
 \pi : \B^2 \rightarrow \{\B_\pi^{(0)}, \B_\pi^{(1)} \}
\end{equation}
that partitions the vector channel $\B^2$, i.e., two independent and identical instances of $\B$, into two bit channels
\begin{align}
 \B_\pi^{(0)}    &: \{0,1\} \rightarrow \mathcal Y^2 \\\nonumber
 \B_\pi^{(1)}    &: \{0,1\} \rightarrow \mathcal Y^2 \times \{0,1\}
\end{align}
with symmetric capacities
\begin{align}
 I(\B_\pi^{(0)}) &= I(U_0 ; Y_0,Y_1) \\\nonumber
 I(\B_\pi^{(1)}) &= I(U_1 ; Y_0,Y_1 | U_0) \; .
\end{align}
The labeling rule is given by
\begin{equation}
\labeling_\pi : \ \ve{u} = [u_0,u_1] \in \F_2^2 \mapsto \ve{u} \cdot \ve{G}_2 \in \F_2^2
\end{equation}
as visualized in Fig.~\ref{Polar_2}. 
Since the average capacity per binary symbol does not change under an SBP, we will denote the mean value of the bit channel capacities of $\pi$ by $I(\B)$ instead of $\mean_\pi(\B)$ in the following. 

It follows easily from~\cite{Arikan:09} by comparison of the permutation matrices that the construction of a polar code of length $N=2^n$ may be equivalently represented by the $n$-fold product concatenation of $\pi$ as defined in the preceding section. 
The resulting SBP $\pi^n$ generates a partition of the vector channel $\B^N$
\begin{equation}
 \pi^n : \B^N \rightarrow \{\B_{\pi^N}^{(0)}, \ldots, \B_{\pi^N}^{(N-1)} \}
\end{equation}
into $N$ bit channels
\begin{equation}
 \B_{\pi^n}^{(i)}    : \{0,1\} \rightarrow \mathcal Y^N \times \{0,1\}^i
\end{equation}
($0\leq i< N$) with symmetric capacities
\begin{equation}
 I(\B_{\pi^n}^{(i)}) := I(U_i ; Y_0,\ldots ,Y_{N-1} | U_0 , \ldots , U_{i-1}) \; .
\end{equation}
Here, the labeling rule is given by
\begin{equation}
\labeling_{\pi^n} : \ \ve{u} \in \F_2^N \mapsto \ve{u} \cdot \ve{G}_N \in \F_2^N \; .
\end{equation}

Therefore, the transmission of each source symbol $u_i$ can be described by its own bit channel $\B_{\pi^n}^{(i)}$.
The output of each channel $\B_{\pi^n}^{(i)}$ depends on the values of the symbols of lower indices $u_0,\ldots u_{i-1}$. 
Thus, the channels $\B_{\pi^n}^{(i)}$ imply a specific decoding order.

For data transmission only the bit channels with highest capacity
are used, referred to as \emph{information channels}. 
The data transmitted over the remaining bit channels (so-called \emph{frozen channels})
are fixed values known to the decoder. 
By this means, the code rate can be chosen in very small steps of $1/N$ without the need for changing the code construction -- 
a property especially useful for polar-coded modulation (cf., Sec.~\ref{sec:polar_mlc}-B).

In order to select the optimal set of frozen channels, the values of
the capacities $I(\B_{\pi^n}^{(i)})$ are required. These 
can either be obtained by simulation or by density evolution~\cite{Mori:09}.

\subsection{Successive Decoding}
Upon receiving a vector $\ve{y}$ --  being a noisy version of the codeword $\ve{c}$ resulting from transmission over the channel $\B$ -- 
the information symbols $u_i$ can be estimated successively for $i=0,\ldots, N-1$. 
Here, information combining~\cite{LandH:2006} of reliability values obtained from the channel output $\ve{y}$ is performed instead of $\F_2$ arithmetics as in the encoding process. 

The successive cancellation (SC) decoding algorithm~\cite{Arikan:09} for polar codes generates estimates on the information symbols $\hat u_i$ (transmitted over the channel $\B_{\pi^n}^{(i)}$) 
one after another, making use of the already decoded symbols $\hat u_0,\ldots,\hat u_{i-1}$. 
We denote the probability that an erroneous decision is made at index $i$ given the previous decisions have been correct, by $p_{\mathrm{e}}(\B_{\pi^n}^{(i)})$. 
Thus, the word error rate for SC decoding ($\mathrm{WER}_{\mathrm{SC}}$) is given by
\begin{equation}\label{sc_wer}
 \mathrm{WER}_{\mathrm{SC}} = 1 - \prod_{i \in \mathcal A}\left(1 - p_{\mathrm{e}}(\B_{\pi^n}^{(i)})\right)
\end{equation}
where $\mathcal A$ denotes the set of indices of the information channels. 

\subsection{Variance of the Bit Channel Capacities}
With increasing block length, the set of bit channels $\B_{\pi^n}^{(i)}$ shows a polarization
effect in the sense that the capacity $I(\B_{\pi^n}^{(i)})$ of almost each bit channel is
either near $0$ or near $1$. The fraction of bit channels not being either completely noisy or
completely noiseless tends to zero \cite{Arikan:09}. 

In the following, we show that 
this polarization effect may be represented by the sequence of variances of the respective polar codes' bit channel capacities for increasing block length.
The variance of the bit channel capacities of a length-$N$ polar code around their mean value $I(\B)$ is given by
\begin{equation}
 V_{\pi^n}(\B^N) = \frac{1}{N}\sum_{i=0}^{N-1} I(\B_{\pi^n}^{(i)})^2 - I(\B)^2 \; .
\end{equation}
Using \eqref{var_conc}, we notice that the sequence of variances increases monotonously as the block length gets larger, i.e.,
\begin{equation}
 V_{\pi^{n+1}}(\B^{2N}) \geq V_{\pi^n}(\B^N) \; .
\end{equation}
Furthermore, from \eqref{max_var} the sequence $\{V_{\pi^n}(\B^N)\}_{n \in \mathbb N}$ is upper-bounded by
\begin{equation}
 V_{\pi^n}(\B^N) \leq I(\B)(1-I(\B))
\end{equation}
for all $n \in \mathbb N$. According to \eqref{max_var}, this maximum variance can be only achieved iff all bit channel capacities $I(\B_{\pi^n}^{(i)})$ are either $0$ or $1$, which 
obviously corresponds to the state of perfect polarization. As shown by Ar{\i}kan~\cite{Arikan:09}, the latter is asymptotically approached while the block length $N$ goes to infinity; 
therefore, we have
\begin{equation}
 \lim_{n \rightarrow \infty} V_{\pi^n}(\B^N) =  I(\B) \cdot (1-I(\B)) \; .
\end{equation}
Although we have not yet been able to establish an explicit relation between bit channel capacity variance and code error performance, one would intuitively expect 
that increasing the variance by a careful code design should correspond to a sharper polarization of the bit channels and therefore should lead to better performing polar 
codes in terms of word error rate or bit error rate.

\begin{figure}
\psfrag{Clabel}[Bc][Bc][0.8]{$I(\B)$}
\psfrag{Vlabel}[Bc][Bc][0.8]{$V_{\pi^n}(\B^N)$}
\centerline{\includegraphics[width=0.48\textwidth]{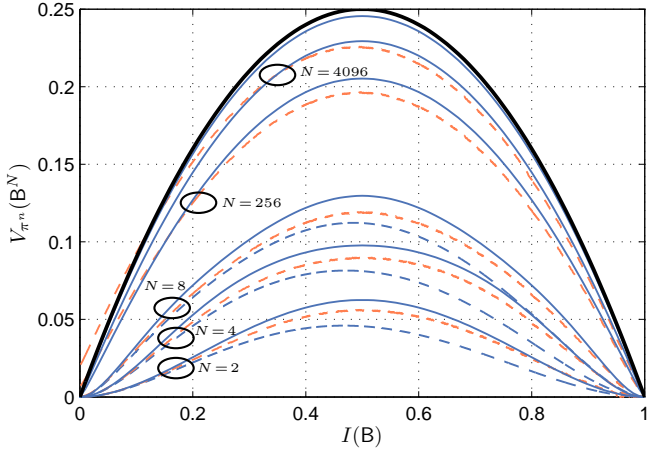}\psellipse(-6.4,1.1)(.25,.15)\psellipse(-6.4,1.5)(.25,.15)\psellipse(-6.45,1.9)(.25,.15)\psellipse(-6.1,3.3)(.25,.15)%
  \psellipse(-5.05,5)(.25,.15)\rput(-5.84,1.1){\tiny $N\!=\!2$}\rput(-5.88,1.6){\tiny $N\!=\!4$}\rput(-6.55,2.2){\tiny $N\!=\!8$}\rput(-5.4,3.3){\tiny $N\!=\!256$}\rput(-4.3,5.05){\tiny $N\!=\!4096$}}
\caption{\label{var_bec}Bit channel variance for polar codes over various B-DMCs, block length $N=2^n$, $n=1,2,3,8,12,20$. Blue solid: BEC, blue dashed: BSC, red dashed: binary-input AWGN channel (Gaussian approximation). 
Black: upper bound on the variance.}
\end{figure}

Fig.~\ref{var_bec} depicts the variance of the bit channel capacities for polar codes of various block lengths constructed over several B-DMCs 
as a function of their capacity. Besides the binary erasure channel (BEC) and the binary symmetric channel (BSC)~\cite{CoverT:91} 
-- that represent the extremes of information combining~\cite{LandH:2006} and serve as an upper and lower bound, resepectively -- 
values for the binary-input AWGN channel are given that have been obtained by density evolution with a Gaussian approximation, as is explained 
in Sec.~\ref{sec:sim}. Obviously, the inaccuracy introduced by this approximation increases with decreasing channel capacity $I(\B)$. 
The converging behaviour for increasing block length $N$ towards the maximum achievable variance (black line) 
can clearly be observed.


\section{Multilevel Polar Coding}
\label{sec:polar_mlc}
\noindent
We now consider the conventional discrete-time equivalent system model of $M$-ary digital pulse-amplitude modulation (PAM)~\cite{Proakis:2000}  -- $M=2^m$ being a power of $2$ --
with signal constellations of real-valued signal points (ASK) or of complex-valued signal points (PSK, QAM etc.) over 
a memoryless channel $\W$, e.g., the AWGN channel. 

From an information-theoretic point of view, an optimal combination of binary coding and $M$-ary modulation follows the multilevel coding (MLC) principle \cite{ImaiH:1977, WachsmannFH:99}. 

\subsection{Multilevel Coding}
In the MLC approach, the $M$-ary channel $\W$ is partitioned into $m$ bit channels (also called \emph{bit levels}) by means of an $m$-SBP
\begin{equation}\label{mlc_labeling}
 \lambda : \W \rightarrow  \{\B_\lambda^{(0)},\ldots, \B_\lambda^{(m-1)} \} \; .
\end{equation}
The mapping from binary labels to amplitude coefficients is specified by the labeling rule $\labeling_\lambda$.

Channel coding is implemented in the MLC setup by using binary \emph{component codes} \cite{WachsmannFH:99} for each of the bit levels $\B_\lambda^{(i)}$ individually 
with correspondingly chosen code rates $R_i$. The overall rate (bit per transmission symbol) is given as the sum $R=\sum_{i=0}^{m-1}R_i$. 
The receiver then performs multi-stage decoding (MSD), i.e., it computes reliability information for decoding of the first bit level which are passed to the decoder of 
the first component code. The decoding results are used for demapping and decoding of the next bit level, and so on.

According to the capacity rule~\cite{WachsmannFH:99}, the code rate for the $i$-th level should match the bit level capacity $I(\B_\lambda^{(i)})$. Since these capacities vary significantly for the different levels, for MLC channel codes are preferred, that allow for a very flexible choice of the code rate. 

The mutual information between the channel input and channel output of $\W$ assuming equiprobable source symbols is also referred to as the 
coded modulation~\cite{WachsmannFH:99}, or constellation-constraint, capacity $C_{\mathrm{cm}}(\W)$. It is related to the average capacity per binary symbol \eqref{def_mean} 
of $\W$  by
\begin{align}\label{eq:cm_cap}
   C_{\mathrm{cm}}(\W)  :=  I(X;Y) =  \sum_{i=0}^{m-1}I(\B_\lambda^{(i)}) = m \cdot \mean_\lambda(\W) \; .
\end{align}
Since $\lambda$ is an SBP, the coded modulation capacity does not depend on the specific labeling rule $\labeling_\lambda$. 

A potential drawback of the MLC approach for practical use lies in the necessity for using several (comparatively short) component codes with varying code rates for the particular bit levels. 

\subsection{Multilevel Polar Coding}
We have shown that both, the multilevel coding construction and the polar coding transform, may be described by SBPs. 
This allows us to represent the combination of MLC with polar codes in a simple form as a product concatenation of SBPs. It also provides insight how the 
labeling $\labeling_\lambda$ should be chosen in an optimal way.

A \emph{multilevel polar code} of length $mN$, i.e., a multilevel code with length-$N$ component polar codes over an $M$-ary constellation, is obtained by the 
order-$mN$ concatenation of the $m$-SBP $\lambda$ of MLC and the $N$-SBP $\pi^n$ of the polar code: 
\begin{equation}
 \lambda \conc \pi^n: \W^{N} \rightarrow \{\B_{\lambda \conc \pi^n}^{(0)},\ldots, \B_{\lambda \conc \pi^n}^{(mN-1)} \}
\end{equation}
as defined in \eqref{def_SBP_conc}. 
The encoding process for this multilevel polar code is described by the generator matrix
\begin{equation}
 \ve{P}_{m,N} \cdot \left( \ve{G}_N \otimes \ve{I}_m \right)
\end{equation}
with $\ve{P}_{m,N}$ as in \eqref{SBP_conc_lin}, followed by labeling and mapping to the $N$ transmit symbols as defined by $\lambda$. Here, $\ve{I}_m$ denotes the $(m,m)$ identity matrix. 
\begin{figure}
\psfrag{chindex}[Bc][Bc][0.8]{bit-channel index $i$ $\rightarrow$}
\psfrag{Cap}[Bc][Bc][0.8]{$I(\B^{(i)})$}
\psfrag{MLClevel0}[Bc][Bc][0.6]{MLC level 0}
\psfrag{MLClevel1}[Bc][Bc][0.6]{MLC level 1}
\psfrag{MLClevel2}[Bc][Bc][0.6]{MLC level 2}
\psfrag{MLClevel3}[Bc][Bc][0.6]{MLC level 3}
\centerline{\includegraphics[width=0.48\textwidth]{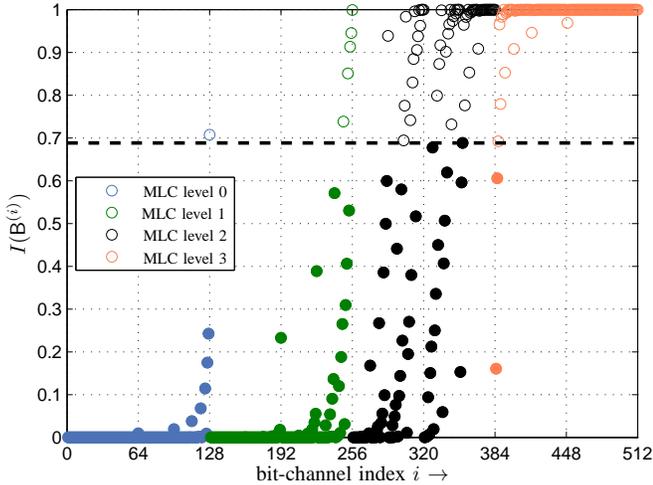}}
\caption{\label{16ask_bitcap}Bit channel capacities of a multilevel polar code ($mN=512$) using $16$-ASK modulation with labeling according to the set-partitioning rule~\cite{Ungerboeck:87} 
over the AWGN channel at $10 \log_{10}(\Es/\No) = 7\,\dB$. 
The overall rate is $R=1.5$. Frozen channels are demarked by filled circles.}
\end{figure}

The word error rate for successive decoding of a multilevel polar code ($\mathrm{WER}_{\mathrm{SC}}$) is given by
\begin{equation}\label{sc_wer_mlc}
 \mathrm{WER}_{\mathrm{SC}} = 1 - \prod_{i \in \mathcal A}\left(1 - p_{\mathrm{e}}(\B_{\lambda \conc \pi^n}^{(i)})\right) \; ,
\end{equation}
in analogy to \eqref{sc_wer}. 
Here, $\mathcal A$ denotes the set of indices of the channels used for information transmission while $p_{\mathrm{e}}(\B_{\lambda \conc \pi^n}^{(i)})$ stands for 
the probability of a first wrong decision at index $i$ in the successive decoding process, like before.  

We remark that the selection of frozen channels -- and thus, the rate allocation -- is done in exactly the same way as for a usual binary polar code by determining the symmetric capacities 
$I(\B_{\lambda \conc \pi^n}^{(i)})$ ($0\leq i < mN$) and choosing the most reliable bit channels for data transmission. This selection process is exemplarily visualized in Fig.~\ref{16ask_bitcap} for an artificial choice of 
parameters. 
Therefore, the explicit application of a rate allocation rule to the particular component codes 
-- like considered in the original MLC approach~\cite{WachsmannFH:99} -- is not needed in case of multilevel polar codes. 
However, it has been shown~\cite{LIT_scc13_seidl_A} that the rate allocations obtained by this method basically equal those obtained from the capacity rule. 

According to \eqref{var_conc}, the variance of the bit channels of a multilevel polar code with length-$N$ component codes is given by
\begin{equation}\label{var_mlc_polar}
 V_{\lambda\conc\pi^n}(\W^N) = V_\lambda(\W) + \frac{1}{m}\sum_{i=0}^{m-1}V_{\pi^n}(\B_\lambda^{(i)}) \; .
\end{equation}
Thus, the SBP $\lambda$ -- that represents the modulation step -- may be seen as the first polarization step of a multilevel polar code. 
From this representation, it is clear that $\lambda$ should be chosen such that it maximizes the term \eqref{var_mlc_polar}. 

In this approach, both binary coding and $2^m$-ary modulation are represented in a unified form as a sequential binary channel partition of the vector channel $\W^N$. Both 
should be designed according to the polarization principle, i.e. the maximization of the variance of the bit channel capacities \eqref{var_mlc_polar} under successive cancellation 
-- or, equivalently, multi-stage -- decoding by careful choice of the labeling rule.

\subsection{Multilevel Polar Codes are Capacity-Achieving}
Using MLC with multi-stage decoding, an $M$-ary channel $\W$ is splitted into $m$ bit levels $\B_\lambda^{(i)}$ ($0\leq i < m$) that are 
B-DMCs as long as $\W$ is a DMC. 
Their symmmetric capacities sum up to $C_{\mathrm{cm}}(\W)$, cf.~\eqref{eq:cm_cap}. 
According to~\cite[Th.~1]{Arikan:09}, the polar component codes approach each of these bit level capacities while their block length increases.

We thus conclude, that multilevel polar codes together with MSD and SC decoding achieve the coded modulation capacity $C_{\mathrm{cm}}(\W)$ for arbitrary $M$-ary 
signal constellations in case of a memoryless transmission channel. All results on the speed of convergence considering transmission over a 
single B-DMC hold as well in the case of MLC.

Obviously, this (asymptotic) result does not depend on the labeling rule $\labeling_\lambda$ applied in MLC. 
However, for finite-length codes the labeling has significant impact on the performance of polar-coded MLC.

\subsection{Influence of the Labeling Rule}
\begin{figure}
\psfrag{Clabel}[Bc][Bc][0.8]{$\mean_\lambda(\W),\mean_{\bar\lambda}(\W)$}
\psfrag{Vlabel}[Bc][Bc][0.8]{$V_\lambda(\W),V_{\bar\lambda}(\W)$}
\centerline{\includegraphics[width=0.48\textwidth]{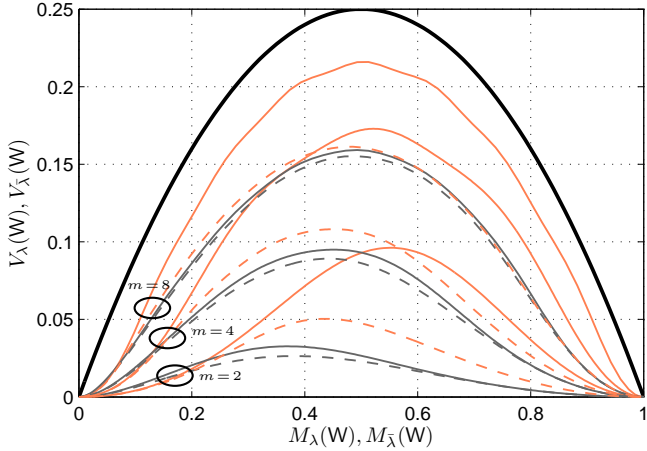}\psellipse(-6.4,1)(.25,.15)\psellipse(-6.5,1.5)(.25,.15)\psellipse(-6.7,1.9)(.25,.15)%
  \rput(-5.8,1){\tiny $m\!=\!2$}\rput(-5.9,1.6){\tiny $m\!=\!4$}\rput(-6.75,2.2){\tiny $m\!=\!8$}}
\caption{\label{var_ask}Bit level variance for $2^m$-ary ASK signalling over the AWGN channel ($m = 2,4,8$). Solid lines: multi-stage decoding, dashed lines: parallel decoding. 
 Red: SP labeling, black: Gray labeling}
\end{figure}

From \eqref{var_mlc_polar}, it is clear that a labeling rule $\labeling_\lambda$ should be applied that leads to a large variance of the bit level capacities. 
Here, we focus on two labeling approaches that follow contrary aims:
\begin{itemize}
  \item In the set-partitioning (SP) labeling approach (corresponding to $\lambda_{\mathrm{SP}}$) by Ungerboeck~\cite{Ungerboeck:87}, for each of the bit levels -- starting from the lowest one -- the sets of signal points corresponding to the 
        following bit level are chosen such that the minimum Euclidean distance within the subsets is maximized. Therefore, the increment of mutual information from each level to the next one 
        is designed to be large -- if there is knowledge about the previous levels -- which should lead to widely separated bit level capacities corresponding to large values of the 
        variance $V_{\lambda_\mathrm{SP}}(\W)$. 
  \item As opposed to that, the Gray labeling approach $\lambda_{\mathrm{G}}$ aims to generate bit levels that are as independent as possible~\cite{StierstorferF:vtc07}. 
        Here, we expect bit levels with capacities that do not differ significantly, 
        leading to a small variance $V_{\lambda_\mathrm{G}}(\W)$ of the bit level capacities.
\end{itemize}
Fig. \ref{var_ask} depicts the variance of the bit levels for ASK modulation using both SP and (binary-reflected) Gray labeling. 
Here, we focus on the variance curves for multi-stage decoding (solid lines); the variances under parallel decoding will be considered in the following section. 
It can be observed that -- except for small capacities $\mean_\lambda(\W)$ -- the SP labeling approach leads to significantly larger bit level variances compared to Gray labeling, as expected. 
Therefore, for multilevel polar codes, SP labeling should be preferably applied.
 
Furthermore, when compared to the corresponding variance curves of polar codes over a single B-DMC for $N=2,4,8$ as shown in Fig.~\ref{var_bec},
especially in case of SP labeling the achieved bit level variance is significantly higher, underlining the importance of the careful choice of the labeling $\labeling_\lambda$ in this 
first step of polarization for multilevel polar codes. 


\section{Bit-Interleaved Polar-Coded Modulation}
\label{sec:polar_bicm}
\noindent
In contrast to the successive approach used for MLC with MSD, in a BICM setup all bit levels are treated equally at both sides, the transmitter and the 
receiver~\cite{CaireTB:98,FabregasMC:now2009}. 

\subsection{Bit-Interleaved Coded Modulation}
We assume here the same underlying $2^m$-ary channel $\W$ as before. 
The source bits in BICM are encoded using a single binary channel code with rate $R_{\mathrm{c}}$, leading to an overall rate of $R=m\cdot R_{\mathrm{c}}$.
The code symbols are (possibly) interleaved according to some pseudo-random order and partitioned into $m$-tuples of code symbols, which are then mapped to amplitude coefficients $x\in \mathcal X$. 

The BICM receiver performs parallel decoding, i.e., it neglects the relations between the bit levels and computes reliability information independently for each bit level based on the received symbol.
These bit metrics are deinterleaved and fed to the decoder. 
Thus, the channel transform used in the BICM setup may be represented by an $m$-PBP $\bar\lambda$
\begin{equation}\label{bicm_lambda}
 \bar \lambda : \W \rightarrow \{\B_{\bar \lambda}^{(0)},\ldots, \B_{\bar \lambda}^{(m-1)} \} \; .
\end{equation}

The BICM capacity\footnote{assuming equiprobable input symbols} (or parallel-decoding capacity) of the channel $\W$ is given as the sum of the 
bit level capacities $I(\B_{\bar\lambda}^{(i)})$ neglecting the feedback of lower bit levels; 
therefore, from \eqref{eq:PDvsMSD} it is generally smaller than the coded-modulation capacity:
\begin{equation}\label{eq:Cbicm}
   C_{\lambda,\mathrm{bicm}}(\W) = \sum_{i=0}^{m-1}I(\B_{\bar\lambda}^{(i)}) \leq C_{\mathrm{cm}}(\W)  \; .
\end{equation}

This loss of the BICM capacity w.r.t. to the coded-modulation capacity depends on $\W$, but also on the applied labeling rule $\labeling_\lambda$. 
It has been shown that -- except for the case of very low capacities $C_{\lambda,\mathrm{bicm}}(\W)$ -- this loss is minimized when Gray labeling is used whereas SP labeling leads to a significant loss of mutual information~\cite{StierstorferF:vtc07,Stierstorfer:diss}. 
The labeling-dependent different behaviour under parallel decoding -- when compared to MSD -- is also evident from the bit level variances as in Fig.~\ref{var_ask}: 
While the curves for MSD and parallel decoding do not differ significantly in case of Gray labeling, for SP labeling with parallel decoding a serious degradation is observed. 
Therefore, we will only consider Gray labeling $\labeling_{\lambda_{\mathrm{G}}}$ in the BICM setup.

\subsection{Bit-Interleaved Polar-Coded Modulation}
\begin{figure}
\psfrag{galp}[Bc][Bc][0.7]{$\oplus$}
\psfrag{bitl3}[Bc][Bc][0.9]{$\labeling_{\lambda_\mathrm{G}}$}
\psfrag{sym0}[Bc][Bc][0.8]{$x_0$}
\psfrag{sym1}[Bc][Bc][0.8]{$x_1$}
\psfrag{u00}[Bc][Bc][0.8]{$u_0$}
\psfrag{u10}[Bc][Bc][0.8]{$u_4$}
\psfrag{u20}[Bc][Bc][0.8]{$u_2$}
\psfrag{u30}[Bc][Bc][0.8]{$u_6$}
\psfrag{u01}[Bc][Bc][0.8]{$u_1$}
\psfrag{u11}[Bc][Bc][0.8]{$u_5$}
\psfrag{u21}[Bc][Bc][0.8]{$u_3$}
\psfrag{u31}[Bc][Bc][0.8]{$u_7$}
\centerline{\includegraphics[width=0.4\textwidth]{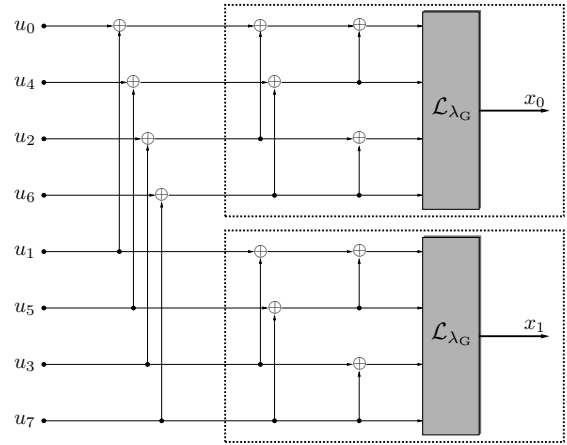}}
\caption{\label{pic_polar8_bicm}Encoding graph for a BICM polar code of length $N=8$ with generator matrix $\ve{G}_8 = \ve{B}_8 \ve{F}_8$ for a $16$-ary constellation. The bit-reversal permutation $\ve{B}_8$ has been already applied to $\ve{u}$.} 
\end{figure}
Since the labeling $\labeling_{\lambda_\mathrm{G}}$ is fixed, there remain two ways for optimizing the combination of polar codes and BICM:
either by designing an optimized interleaver or by changing the polar code itself. 

In~\cite{ShinLimYang:12}, the interleaver design has been considered. 
Clearly, the bit channel variance for a length-$mN$ BICM polar code depends on how the bit channels $\B_{\bar\lambda_\mathrm{G}}^{(i)}$ obtained from the $N$ transmission symbols 
-- with varying capacities -- are allocated to the 
order-$mN$ polar coding transform. The authors showed that by means of a partial exhaustive search a performance improvement can be observed when compared to random interleaving~\cite{ShinLimYang:12}. 

Here, we will follow the second approach: We assume that no interleaver is used at all. 
Since we focus on memoryless transmission channels such as the AWGN channel in this work, this is a reasonable assumption.\footnote{We are motivated by the fact that 
for BICM with convolutional codes over the AWGN channel, even a significant performance gain for the interleaver-free case w.r.t. random interleaving can be observed~\cite{LIT_izs_2010_cst}.} 
Now, the straight-forward approach of combining BICM over an $2^m$-ary constellation with polar codes simply connects a polar code 
of length $mN$ -- described by a generator matrix 
$\ve{G}_{mN}$ -- to the $m$-PBP $\bar\lambda_{\mathrm{G}}$. In order to use Ar{\i}kan's standard construction, we assume $m$ to be a power of two itself. 
Otherwise, we would have to use a polar code with a different kernel.
Fig.~\ref{pic_polar8_bicm} shows an example of a simple BICM polar code obtained in this way 
where the input symbols of a length-$8$ polar code are mapped onto two symbols of a $16$-ary constellation. 

The overall channel transform for this unpermuted approach is given by $\bar\lambda_{\mathrm{G}} \concb \pi^{\log_2(m)} \conc \pi^n$. From Sec.~\ref{sec:ch_transf}-D 
we know that the first part 
$\bar\lambda_{\mathrm{G}} \concb \pi^{\log_2(m)}$ may be seen as a degraded $m$-SBP, represented by the labeling rule $\labeling_{\lambda_{\mathrm{G}}\concb\pi^{\log_2(m)}}$. 
Since $\labeling_{\lambda_{\mathrm{G}}}$ is fixed, our optimization approach for polar-coded BICM consists in changing the first polarization steps of the polar code, i.e., 
we replace the $m$-SBP $\pi^{\log_2(m)}$ by 
an optimized $m$-SBP $\tau$ that maximizes the bit channel variance of $\bar\lambda_{\mathrm{G}}\concb \tau$.

\subsection{Transformation of Labelings}
It has been shown~\cite{Alvarado:12} that for one-dimensional constellations, natural labeling and binary reflected Gray labeling can be transformed into each other by a (bijective) 
linear transform. 
\subsubsection{ASK/PSK Constellations}
A natural labeling (counting in dual numbers) over an $M$-ary ASK/PSK constellation -- which is identical to an SP labeling in this case -- can be represented as an 
$(M, m)$ binary matrix $\ve{M}_{\mathrm{SP},m}$ 
($m = \log_2(M)$) containing the dual representations of the 
numbers $0,\ldots,M-1$ as rows. Here, the left-most column represents the least significant bit. Similarly, a (binary reflected) Gray labeling is given 
by a binary matrix $\ve{M}_{\mathrm{Gray}}$ of equal dimensions. 
Below, an example for $m=3$ is given:
\begin{equation} \ve{M}_{\mathrm{SP},3}\! = \begin{bmatrix}0 & 0 & 0 \\
                                                           1 & 0 & 0 \\
                                                           0 & 1 & 0 \\
                                                           1 & 1 & 0 \\
                                                           0 & 0 & 1 \\
                                                           1 & 0 & 1 \\
                                                           0 & 1 & 1 \\
                                                           1 & 1 & 1 \end{bmatrix}\; , \quad
                \ve{M}_{\mathrm{Gray},3}\!= \begin{bmatrix}0 & 0 & 0 \\
                                                         1 & 0 & 0 \\
                                                         1 & 1 & 0 \\
                                                         0 & 1 & 0 \\
                                                         0 & 1 & 1 \\
                                                         1 & 1 & 1 \\
                                                         1 & 0 & 1 \\
                                                         0 & 0 & 1 \end{bmatrix}\; .
\end{equation}
As shown in \cite{Alvarado:12}, a set-partitioning labeling of an $M$-ASK/PSK constellation can be transformed into a binary reflected Gray labeling via an $(m , m)$ binary matrix
\begin{equation}\label{matrix:sp2gray}\ve{T}_m = \begin{bmatrix}1 & 0 & 0 & \ldots & 0 & 0\\
                            1 & 1 & 0 & \ldots & 0 & 0\\
                            0 & 1 & 1 & \ldots & 0 & 0\\
                            \mbox{} & \vdots & \mbox{} & \ddots & \vdots & \vdots\\
                            0 & 0 & 0 & \ldots & 1 & 1 \end{bmatrix}
\end{equation}
such that 
\begin{equation}
 \ve{M}_{\mathrm{SP},m} \cdot \ve{T}_m = \ve{M}_{\mathrm{Gray},m}
\end{equation}
holds.

\subsubsection{QAM Constellations}
Similar to the case of ASK/PSK constellations, it is also possible to convert an SP labeling into a Gray labeling by a linear transform in case of square $M^2$-QAM constellations.
Here, $\ve{M}_{\mathrm{SP},2m}$ and $\ve{M}_{\mathrm{Gray},2m}$ are related by
\begin{equation}\label{qam_trans}
 \ve{M}_{\mathrm{SP},2m} \cdot (\ve{G}_2 \otimes \ve{T}_m) = \ve{M}_{\mathrm{Gray},2m}
\end{equation}
where $\ve{G}_2$ equals the generator matrix of a length-$2$ polar code, cf.,~\eqref{eq:polarc:G}. 
This relation is proven in Appendix B. 

\subsubsection{Successive Decoding of $\ve{T}_m$}
Since $\ve{T}_m$ is a non-singular, square binary $(m,m)$ matrix, it induces a channel transform which is represented by the $m$-SBP
\begin{equation}\label{def_tau}
 \tau : (\B_0 \times \ldots \times \B_{m-1}) \rightarrow \{\B_{\tau}^{(0)},\ldots, \B_{\tau}^{(m-1)} \} \; .
\end{equation}
that maps the vector channel of $m$ independent B-DMCs $\B_i$ ($0\leq i < m$) to an ordered set of different B-DMCs. 
By construction of $\tau$, the concatenation $\bar\lambda_{\mathrm{G}} \concb \tau$ is characterized by a labeling rule $\labeling_{\lambda_{\mathrm{G}}\concb\tau} = \labeling_{\lambda_{\mathrm{SP}}}$, 
i.e., an SP labeling.

We now demonstrate that the transform induced by the matrix $\ve{T}_m$ can be reversed in a successive way at the receiver side just like the polar coding transform $\pi^n$ -- 
that is induced by $\ve{G}_N$ -- under SC decoding. 
Let $\ve{x} = \ve{u}\ve{T}_m$ be the Gray-labeled representation of $\ve{u} = [u_0,\ldots ,u_{m-1}]$ that is mapped to $M$-ary ASK/PSK symbols and transmitted. 
As follows immediately from the structure of $\ve{T}_m$~\eqref{matrix:sp2gray}, $\ve{x}$ is given as
\begin{equation}\label{map_tm}
 \ve{x} = [ u_0 \oplus u_1, u_1 \oplus u_2,\ldots , u_{m-2} \oplus u_{m-1}, u_{m-1} ]\; .
\end{equation}
Let us further assume that, at the receiver, reliability values $L(x_0),\ldots ,L(x_{m-1})$ for the components of $\ve{x}$, e.g., LLR values, have been determined by using 
parallel decoding, like in plain BICM.
\begin{figure}
\psfrag{Lu1}[Bc][Bc][0.6]{$L(u_0)$}
\psfrag{Lu2}[Bc][Bc][0.6]{$L(u_1)$}
\psfrag{Lu3}[Bc][Bc][0.6]{$L(u_2)$}
\psfrag{Lu4}[Bc][Bc][0.6]{$L(u_3)$}
\psfrag{Lx1}[Bc][Bc][0.6]{$L(x_0)$}
\psfrag{Lx2}[Bc][Bc][0.6]{$L(x_1)$}
\psfrag{Lx3}[Bc][Bc][0.6]{$L(x_2)$}
\psfrag{Lx4}[Bc][Bc][0.6]{$L(x_3)$}
\psfrag{u1}[Bc][Bc][0.6]{$u_0$}
\psfrag{u2}[Bc][Bc][0.6]{$u_1$}
\psfrag{u3}[Bc][Bc][0.6]{$u_2$}
\psfrag{la}[Bc][Bc][0.9]{a)}
\psfrag{lb}[Bc][Bc][0.9]{b)}
\psfrag{lc}[Bc][Bc][0.9]{c)}
\psfrag{ld}[Bc][Bc][0.9]{d)}
\centerline{\includegraphics[width=0.4\textwidth]{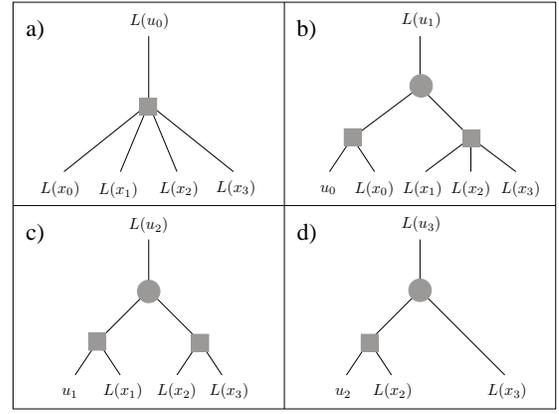}}
\caption{\label{Tm_succ_dec}Decoding trees for successive estimation of $\ve{u} = [u_0,u_1,u_2,u_3]$ from $\ve{x} = \ve{u}\ve{T}_4$ given reliability values $L(x_0), \ldots , L(x_3)$. 
 The known values $u_0, u_1, u_2$ in the graphs correspond to LLR values of $\pm \infty$.}
\end{figure}
The components $u_i$ of $\ve{u}$ ($0\leq i < m$) can now be decoded successively from $\ve{x}$, making use of the reliability information on $\ve{x}$ as well as 
of the already estimated components $u_0,\ldots ,u_{i-1}$:
\begin{itemize}
 \item Clearly, from \eqref{map_tm} $u_0$ may be written as a sum of all components of $\ve{x}$:
\begin{equation}
 u_0 = {\sum_{i=0}^{m-1}}{\kern-1.4em}{\oplus}\hspace*{3mm}x_i \; .\hspace*{37mm}
\end{equation}
In a factor-graph notation, the decoding tree for estimating $u_0$ simply consists of a check node of order $m$, as visualized 
in Fig.~\ref{Tm_succ_dec}a). Therefore, given reliability information on the components of $\ve{x}$, this (Galois field) sum can be evaluated by 
using the well-known operations of information combining, cf., e.g., \cite{LandH:2006}. 
 \item The next component $u_1$ is represented by two independent equations, making use of the knowledge of $u_0$:
\begin{align}
  u_1 &= {\sum_{i=1}^{m-1}}{\kern-1.4em}{\oplus}\hspace*{3mm}x_i \; ,\hspace*{39mm}\\\nonumber
  u_1 &= x_{0} \oplus u_{0} \; .\hspace*{39mm}
\end{align}
Here, ``independent'' means that each code symbol $x_i$ appears in at most one of the equations. The corresponding computation tree is shown in Fig.~\ref{Tm_succ_dec}b), involving two check nodes 
and one variable node.
 \item The remaining components of $\ve{u}$ are now determined one after another in a similar way from the two independent equations
\begin{align}
  u_j &= {\sum_{i=j}^{m-1}}{\kern-1.4em}{\oplus}\hspace*{3mm}x_i \; ,\\\nonumber
  u_j &= x_{j-1} \oplus u_{j-1} \; , \quad j=2,\ldots ,m-1 \; .
\end{align}
\end{itemize}

\begin{figure}
\psfrag{galp}[Bc][Bc][0.7]{$\oplus$}
\psfrag{bitl3}[Bc][Bc][0.9]{$\labeling_{\lambda_\mathrm{G}}$}
\psfrag{sym0}[Bc][Bc][0.8]{$x_0$}
\psfrag{sym1}[Bc][Bc][0.8]{$x_1$}
\psfrag{u00}[Bc][Bc][0.8]{$u_0$}
\psfrag{u10}[Bc][Bc][0.8]{$u_2$}
\psfrag{u20}[Bc][Bc][0.8]{$u_4$}
\psfrag{u30}[Bc][Bc][0.8]{$u_6$}
\psfrag{u01}[Bc][Bc][0.8]{$u_1$}
\psfrag{u11}[Bc][Bc][0.8]{$u_3$}
\psfrag{u21}[Bc][Bc][0.8]{$u_5$}
\psfrag{u31}[Bc][Bc][0.8]{$u_7$}
\centerline{\includegraphics[width=0.4\textwidth]{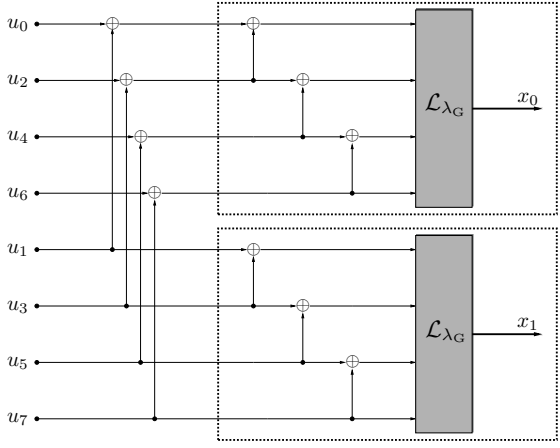}}
\caption{\label{Polar8bicmopt}Encoding graph for an optimized BICM polar code of length $N=8$ with generator matrix $\ve{P}_{4,2}\cdot \left(\ve{G}_2 \otimes \ve{T}_4\right)$ for a $16$-ASK constellation. The permutation $\ve{P}_{4,2}$ has already been applied to $\ve{u}$.}
\end{figure}

\subsection{Code Modification}
Employing $\tau$ in the construction of a length-$mN$ BICM polar code, the overall channel transform is given by 
\begin{equation}\label{opt_bicm}
 (\bar\lambda_\mathrm{G} \concb \tau) \conc \pi^n : \ \W^N \rightarrow  \{\B_{(\bar\lambda_\mathrm{G} \concb \tau) \conc \pi^n}^{(0)},\ldots, \B_{(\bar\lambda_\mathrm{G} \concb \tau) \conc \pi^n}^{(mN-1)} \} \; ;
\end{equation}
thus, the encoding process for this modified BICM polar code is described by a generator matrix
\begin{equation}
 \ve{P}_{m,N} \cdot \left( \ve{G}_N \otimes \ve{T}_m \right) \; ,
\end{equation}
followed by Gray-labeled mapping to the transmit symbols.
Fig.~\ref{Polar8bicmopt} depicts an example of a length-$8$ BICM polar code optimized for $16$-ASK modulation that is described by the generator matrix 
$\ve{P}_{4,2} \cdot \left( \ve{G}_2 \otimes \ve{T}_4 \right)$. 

Interestingly, the transform \eqref{opt_bicm} -- that is optimized for BICM polar codes -- and the optimal multilevel code defined by the $mN$-SBP using SP labeling 
\begin{equation}
 \lambda_\mathrm{SP} \conc \pi^n : \ \W^N \rightarrow  \{\B_{\lambda_{\mathrm{SP}} \conc \pi^n}^{(0)},\ldots, \B_{\lambda_{\mathrm{SP}} \conc \pi^n}^{(mN-1)} \}
\end{equation}
share the same labeling rule and thus decribe the same code, i.e., identical binary source symbols are encoded to identical transmission symbols in both cases. 
However, the decoding strategies at the bit metrics calculation step differ for the two approaches: 
In case of BICM, parallel decoding is used in contrast to successive decoding in the MLC approach.


\section{Simulation Results}
\label{sec:sim}
\noindent
We now give some numerical results in terms of rate-vs.-power-efficiency plots in order to illustrate the error performance of polar-coded modulation with SC decoding over the AWGN channel.

\begin{figure}
\psfrag{ebno}[Bc][Bc][0.8]{$10\log_{10}(E_\mathrm{b}/N_\mathrm{0})$}
\psfrag{rate}[Bc][Bc][0.8]{$R$}
\centerline{\includegraphics[width=0.48\textwidth]{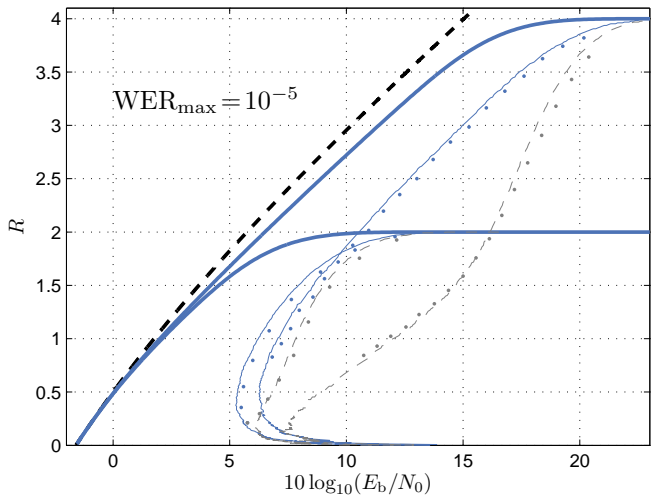}\rput(-6,5.3){$\mathrm{WER}_{\mathrm{max}}\!=\!10^{-5}$}}
\caption{\label{sim_16ask_acc}$M$-ASK / AWGN, $M=4,16$: Rate vs. SNR of multilevel polar codes using SP labeling (blue) and Gray labeling (gray) obtained by DE (continuous lines)
as well as simulated values. 
Overall block length $mN=512$. 
Bold blue line: coded-modulation capacity, dashed black: Shannon bound for real constellations}
\end{figure}

Besides common Monte-Carlo simulations, we also present results obtained by density evolution (DE)~\cite{Mori:09, Urban:08}, a method that allows for approximate error performance analysis with 
neglegible numerical effort even for large code lengths. 
Here, for multilevel polar codes, we numerically determine the bit level capacities $I(\B_\lambda^{(i)})$ ($0\leq i < m$) of the respective PAM constellation, cf., e.g.,~\cite{WachsmannFH:99}. 
Now, for each of the $m$ binary component polar codes, a Gaussian channel with capacity $I(\B_\lambda^{(i)})$ is assumed as a transmission channel. 
The $mN$ bit channel capacities -- and the corresponding error probabilities $p_e(\B_{\lambda \conc \pi^n}^{(i)})$ -- of the $m$ component polar codes are then determined
by performing density evolution (DE) with the well-known Gaussian approximation~\cite{ChungUrbanke:01}, i.e., we simply assume the output bit channels of each SBP in the chain 
$\lambda \conc \pi \conc \ldots \conc \pi$ to be AWGN channels. 
Finally, from \eqref{sc_wer_mlc}, the maximum achievable code rate $R$ under successive decoding given a target word error rate $\mathrm{WER}_{\mathrm{max}}$ is obtained. 
This procedure is carried out for each value of the signal-to-noise ratio $E_\mathrm{b}/N_\mathrm{0}$. 
Although the overall transmission channel is the AWGN channel, for the bit channels occurring in the multi-stage decoding process, this assumption certainly does not hold.
Nevertheless, the inaccuracy induced by this Gaussian assumption is small for multilevel polar codes, as shown in Fig. \ref{sim_16ask_acc}.

\begin{figure}
\psfrag{ebno}[Bc][Bc][0.8]{$10\log_{10}(E_\mathrm{b}/N_\mathrm{0})$}
\psfrag{rate}[Bc][Bc][0.8]{$R$}
\centerline{\includegraphics[width=0.48\textwidth]{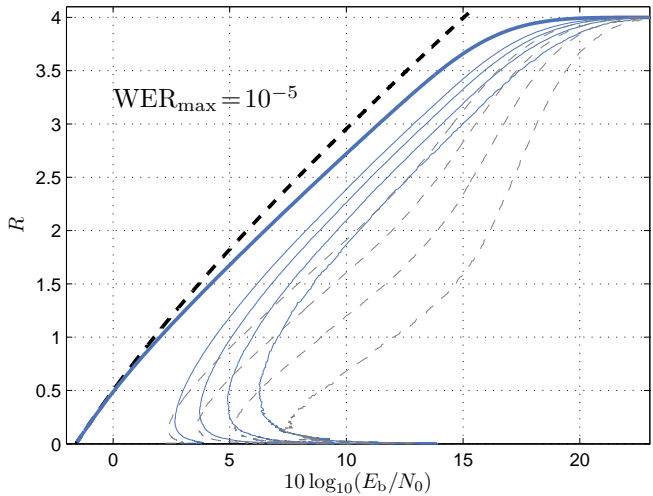}\rput(-6,5.3){$\mathrm{WER}_{\mathrm{max}}\!=\!10^{-5}$}}
\caption{\label{sim_16ask}$16$-ASK / AWGN: Rate vs. SNR of multilevel polar codes using SP (blue) and Gray labeling (dashed gray). 
Overall block length (from right to left) $mN=2^k$, 
$k=9,11,13,15$. Bold blue line: coded-modulation capacity, dashed black: Shannon bound for real constellations}
\end{figure} 
Fig.~\ref{sim_16ask} depicts the performance of multilevel polar codes with $16$-ASK modulation under SC decoding for different labelings $\labeling_\lambda$ and various block lengths. 
The large performance loss of Gray labeling w.r.t. SP labeling can clearly be observed. 

\begin{figure}
\psfrag{ebno}[Bc][Bc][0.8]{$10\log_{10}(E_\mathrm{b}/N_\mathrm{0})$}
\psfrag{rate}[Bc][Bc][0.8]{$R$}
\centerline{\includegraphics[width=0.48\textwidth]{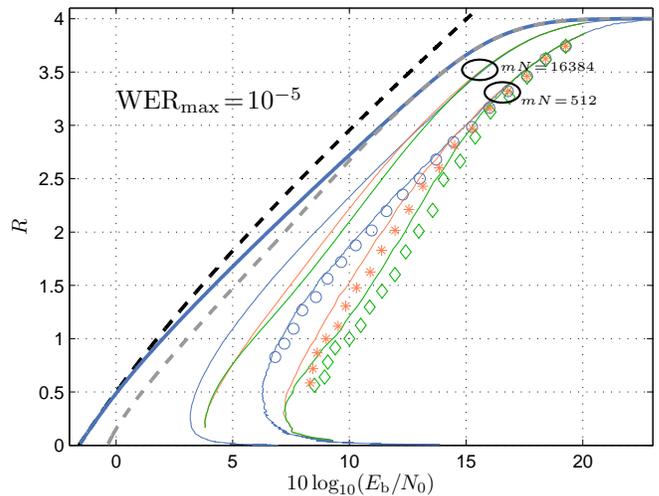}\psellipse(-2.1,5.4)(.25,.15)\psellipse(-2.4,5.7)(.25,.15)%
  \rput(-1.35,5.3){\tiny $mN\!=\!512$}\rput(-1.5,5.75){\tiny $mN\!=\!16384$}\rput(-6,5.3){$\mathrm{WER}_{\mathrm{max}}\!=\!10^{-5}$}}
\caption{\label{sim_16ask_bicm}$16$-ASK / AWGN: Rate vs. SNR of multilevel polar codes using SP labeling (blue) 
and BICM polar codes using the original construction (green) and the proposed modified construction (red). Solid lines correspond to results obtained by DE, markers to simulated values. 
Overall block length (from right to left) $mN=2^k$, 
$k=9,14$. Bold blue line: coded-modulation capacity, dashed gray: BICM capacity using Gray labeling, dashed black: Shannon bound for real constellations}
\end{figure} 
For DE in the case of BICM polar codes, the bit channel capacities from the first polarization steps $I(\B_{\bar\lambda_\mathrm{G}\concb\pi^{\log_2(m)}}^{(i)})$ and 
$I(\B_{\bar\lambda_\mathrm{G}\concb\tau}^{(i)})$ -- as in \eqref{bicm_lambda} and \eqref{def_tau}, 
respectively -- have been obtained by Monte-Carlo simulation, followed by Gaussian-approximated DE for the component codes, as in the MLC case. 
From Fig.~\ref{sim_16ask_bicm}, a significant performance gain for the optimized code construction from Sec.~\ref{sec:polar_bicm}-D w.r.t. unmodified BICM polar codes can be observed. 
However, due to the suboptimality of the BICM approach, the performance of multilevel polar codes is not achieved. 
Moreover, the inaccuracy introduced by the Gaussian assumption for DE increases for the BICM channels when compared to the MLC case, leading to an additional loss. 

\begin{figure}
\psfrag{ebno}[Bc][Bc][0.8]{$10\log_{10}(E_\mathrm{b}/N_\mathrm{0})$}
\psfrag{rate}[Bc][Bc][0.8]{$R$}
\centerline{\includegraphics[width=0.48\textwidth]{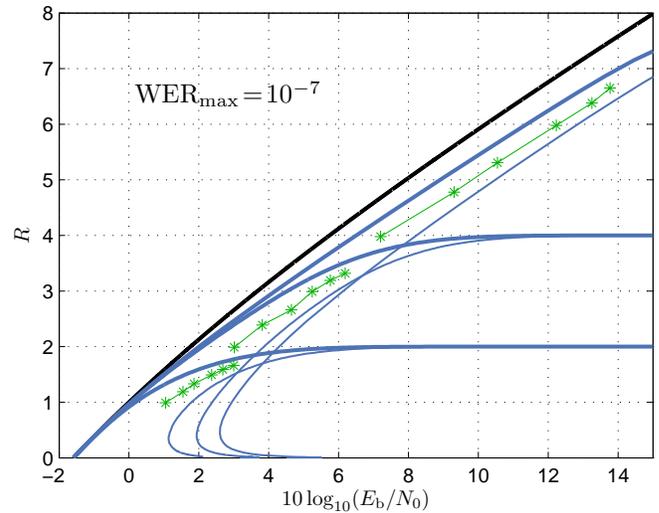}\rput(-5.75,5.6){$\mathrm{WER}_{\mathrm{max}}\!=\!10^{-7}$}}
\caption{\label{sim_qam_comp}$M^2$-QAM / AWGN, $M^2=4,16,256$: Rate vs. SNR of multilevel polar codes over $M^2$-QAM using SP labeling obtained by DE (blue solid lines) 
and reference values for DVB-T2~\cite{DVBT2:12} (green markers). 
Overall block length: $65.536$ (Polar Codes), $64.800$ (LDPC+BCH). 
Bold blue: coded-modulation capacity, black: Shannon bound}
\end{figure} 
Finally, Fig.~\ref{sim_qam_comp} compares the performance of SP-labeled multilevel polar codes to the BICM-based coding scheme used in the DVB-T2 standard~\cite{DVBT2:12}. 
It is observed that multilevel polar codes (under SC decoding) do not achieve the error performance of the DVB-T2 system consisting of a concatenation of an LDPC code with a BCH code of 
equivalent overall block length. 
On the other hand, multilevel polar codes are decoded with a single-step, non-iterative decoding algorithm that requires less information combining operations and thus leads to 
a reduced computational complexity, compared to the concatenated coding approach in DVB-T2.


\section{Conclusions}
\label{sec:conclusions}
\noindent
In this paper, we have extended the binary polar coding approach to higher-order digital $2^m$-ary modulation. 
We have shown that the combination of multilevel coding and polar coding results in a sequential binary channel partition (SBP) of a vector channel into B-DMCs that can be 
successively decoded, just like for the case of binary polar codes. 
The optimal choice of the binary labeling of the $2^m$ signal constellation points has been discussed. 
Using BICM instead of MLC, we have demonstrated -- for the case of ASK, PSK and square QAM constellations -- that by a slight modification of the polar code generator matrix, multilevel polar codes 
and BICM polar codes can be transformed into each other. Although both approaches may be designed to describe the same $2^m$-ary code, for BICM a degradation w.r.t. the multilevel approach 
is observed which is caused by the suboptimal parallel decoding step at the 
bit metrics calculation in BICM. 

Therefore, we conclude that for polar-coded modulation, the use of MLC should be preferred over BICM, if successive decoding is considered. 


\appendices
\section{Proof of Equation \eqref{var_conc}}
By definition, the variance of the bit channel capacities under the concatenation $\varphi \conc \psi$ is given by 
\[V_{\varphi \conc \psi}(\W^{k_2}) = \frac{1}{k_1k_2} \sum_{i=0}^{k_1-1} \sum_{j=0}^{k_2-1} I(\B_{\varphi \conc \psi}^{(k_2i+j)})^2 - \mean_{\varphi \conc \psi}(\W^{k_2})^2 \; .\]
Adding and subtracting the term $\frac{1}{k_1} \sum_{i=0}^{k_1-1} I(\B_{\varphi}^{(i)})^2$ together with \eqref{mean_invar} leads to
\begin{align*}
 V_{\varphi \conc \psi}(\W^{k_2}) &= \frac{1}{k_1} \sum_{i=0}^{k_1-1} I(\B_{\varphi}^{(i)})^2 - \mean_\varphi(\W)^2 \\
                            &\hspace*{2mm}+ \frac{1}{k_1} \sum_{i=0}^{k_1-1} \frac{1}{k_2} \sum_{j=0}^{k_2-1} \left(I(\B_{\varphi \conc \psi}^{(k_2i+j)})^2 - I(\B_\varphi^{(i)})^2\right)  .
\end{align*}
Finally, \eqref{mean_conc} and \eqref{def_variance} yield
\[V_{\varphi \conc \psi}(\W^{k_2}) = V_{\varphi}(\W) + \frac{1}{k_1} \sum_{i=0}^{k_1-1} V_\psi(\B_\varphi^{(i)}) \; .\]

\section{Proof of Equation \eqref{qam_trans}}
We consider a square $M^2$-QAM constellation with labels that are binary tuples of length $2m$ (with $m=\log_2(M)$) of the form
\[\ve{a} := [a_{1,1}, \ldots, a_{1,m}, a_{2,1}, \ldots, a_{2,m}] \]
where the first and last $m$ bits represent the naturally labeled row and column indices, respectively. 
The application of the transform $\ve{G}_2 \otimes \ve{I}_m$ -- with $\ve{I}_m$ being the $(m,m)$ identity matrix -- leads to the following labels 
\[ [(a_{1,1}\oplus a_{2,1}), \ldots, (a_{1,m}\oplus a_{2,m}), a_{2,1}, \ldots, a_{2,m} ]\; ,\]
i.e., the first $m$ bits of each label hold the component-wise modulo-$2$ sum of row and column index. 
It is easily verified that this labeling in fact represents a set-partitioning.

We will show now that this set-partitioned square $M^2$-QAM constellation can be transformed into a Gray-labeled constellation by a simple linear transform, just like for the 
case of ASK/PSK. 

Since the transform $\ve{G}_2 \otimes \ve{I}_m$ is obviously self-inverse, by application to the SP-labeled constellation we obtain again
\[a = [a_{1,1}, \ldots, a_{1,m}, a_{2,1}, \ldots, a_{2,m}] \; .\]
From \eqref{matrix:sp2gray}, the transform $\ve{I}_2 \otimes \ve{T}_m$ applies a (binary reflected) Gray labeling independently to the (now naturally labeled) row and column 
indices which obviously describes a Gray-labeled $M^2$-QAM constellation.

In summary, 
  \[\ve{G}_2 \otimes \ve{T}_m = \left( \ve{G}_2 \otimes \ve{I}_m \right) \left( \ve{I}_2 \otimes \ve{T}_m \right)\]
transforms an SP-labeled $M^2$-QAM constellation into a Gray-labeled one where $\ve{T}_m$ denotes the linear transform from \eqref{matrix:sp2gray}.


\bibliographystyle{IEEEtran}
\bibliography{mjssBib}

%
\end{document}